\documentstyle[12pt,fleqn,epsf]{article}
\newcommand{\pq}{\mbox{$\not\!p$}}
\newcommand{\sq}{\mbox{$\not\!s$}}
\begin{document}
\setlength{\unitlength}{1cm}
\setlength{\mathindent}{0cm}
\thispagestyle{empty}
\null
\hfill WUE-ITP-98-012\\
\null
\hfill UWThPh-1998-14\\
\null
\hfill HEPHY-PUB 691/98\\
\null
\hfill hep-ph/9804306\\
\vskip .8cm
\begin{center}
{\Large \bf Spin Correlations in Production\\[.5em]
and Decay of Charginos%
}
\vskip 2.5em
{\large
{\sc G.~Moortgat-Pick$^{a}$\footnote{e-mail:
    gudi@physik.uni-wuerzburg.de}, 
H. Fraas$^{a}$, A.~Bartl$^{b}$,
 W.~Majerotto$^{c}$}%
}\\[1ex]
{\normalsize \it
$^{a}$ Institut f\"ur Theoretische Physik, Universit\"at
W\"urzburg, Am Hubland, D-97074~W\"urzburg, Germany}\\
{\normalsize \it
$^{b}$ Institut f\"ur Theoretische Physik, Universit\"at Wien, 
Boltzmanngasse 5, A-1090 Wien}\\
{\normalsize \it
$^{c}$ Institut f\"ur Hochenergiephysik, \"Osterreichische
 Akademie der Wissenschaften, Nikolsdorfergasse 18 
, A-1050 Wien}
\vskip 1em
\end{center} \par
\vskip .8cm

\begin{abstract}
We study
 the production of charginos 
$e^+ e^- \to\tilde{\chi}^{+}_i\tilde{\chi}^{-}_j, (i,j=1,2)$ with
 polarized beams and the subsequent  decays $\tilde{\chi}^{+}_{i} \to 
\tilde{\chi}^0_k
\ell^{+}\nu_\ell$ and 
$\tilde{\chi}^{-}_{j} \to \tilde{\chi}^0_l \ell^{-}\bar{\nu}_\ell, 
(k,l=1,\ldots,4)$, including the complete spin 
correlations between production and decay.
Analytical formulae are presented for the 
joint spin-density matrix of the 
charginos, for the chargino decay matrix and for the differential
 cross section of the combined  processes of production and decays.
We present numerical results for pair production of the
 lighter chargino  with unpolarized beams and the leptonic decay of 
$\tilde{\chi}_1^-$
 into the lightest neutralino $\tilde{\chi}^0_1$. The 
lepton angular distribution and the forward-backward asymmetry are studied in four representative scenarios 
 for $\sqrt{s}=192$~GeV and $\sqrt{s}=200$~GeV. 
\end{abstract}
\newpage
\section{Introduction}
\vspace{-.3cm}
The search for supersymmetric particles is one of the main goals of
LEP and of future
$e^{+}e^{-}$ colliders. The charginos, the supersymmetric partners of the charged gauge
and Higgs bosons, are of particular interest as they are expected to be lighter than the
strongly interacting gluino and squarks. The lighter chargino
$\tilde{\chi}^\pm_1$ could first 
be observed in  experiments at $e^{+}e^{-}$ colliders. 

Most studies of chargino production $e^+e^- \to 
\tilde{\chi}^{+}_i 
\tilde{\chi}^{-}_j , i,j = 1,2,$ and chargino decays have been 
performed in the Minimal Supersymmetric Standard Model (MSSM). (See, for 
example 
\cite{bartl, ambrosiano, desy}, and 
references therein.)
 For a clear identification of charginos a precise analysis of their decay characteristics
is indispensable. Angular distributions and angular correlations of the decay
products of the charginos give valuable information on their gaugino
and higgsino components
and thus on the parameters of the MSSM.

Since angular distributions depend on the polarization of the parent
particles one has to take into account the spin correlations between
production and decay of the charginos. In general quantum mechanical
interference effects between various polarization states of the
decaying particles preclude simple factorization of the differential
cross section into a production and a decay factor \cite{tsai,tata}, 
unless the production amplitude is dominated by a single spin
component \cite{feng}. Recently, pair production and decay of the lighter chargino with spin correlations has also been studied in \cite{choi}.
 A variety of event
generators for production and decay of SUSY particles, such as DFGT, SUSYGEN, GRACE and CompHEP 
\cite{dfgt},  have been developed which partly 
also include spin correlations between production and decay. 

In the present paper we study the process 
$e^{+} e^{-} \to \tilde{\chi}_i^{+} \tilde{\chi}_j^{-}$,
 $(i,j=1,2)$ with polarized beams and the subsequent leptonic decays 
$\tilde{\chi}_i^{+} \to \tilde{\chi}_k^0 \ell^{+} \nu_{\ell},
\tilde{\chi}^{-}_j \to \tilde{\chi}^0_l \ell^{-}
\bar{\nu}_{\ell}$.
The main purpose of this paper is the presentation of analytical
formulae for the complete spin correlations between production and decay. The
computation is done for complex couplings so that the formulae are
also useful for the study of CP violating phenomena. The expression
for the differential cross section is composed of the joint spin
density matrix of the charginos and the decay matrices for their
leptonic decay into one of the neutralino states.

The analytical  expressions given for the production density matrix
can also be used as a building block for processes with other chargino
decay channels as e.g. hadronic decay or sequential chargino decay.
 Furthermore, the decay matrix can be combined with other chargino
production processes as e.g.  $e^{-} \gamma$ or $\gamma
\gamma$ collisions.

Since in this article our emphasis is on the analytical formulae we restrict ourselves
in the numerical calculations to the pair production
of the lighter chargino with unpolarized beams, and to the 
leptonic decay of one chargino into the lightest neutralino
$\tilde{\chi}^0_1$.

In sect.~2 we present the analytical expressions for the
spin--density matrix for production, for the decay matrix and for the
differential cross section. Numerical results for the lepton angular
distribution for LEP 2 energies $\sqrt{s}=192$~GeV and
$\sqrt{s}=200$~GeV are given in sect.~3 for four scenarios which
 differ significantly in the mixing character of the chargino and the lightest neutralino.

\section{Analytical formulae}
\vspace{-.3cm} 
\subsection{Definition of couplings}
We study the process 
\begin{equation}
e^{-}(p_1)+e^{+}(p_2) \to \tilde{\chi}^{+}_i(p_3)+ 
\tilde{\chi}^{-}_j(p_4),\label{eq:p1}
\end{equation}
 where the charginos decay leptonically: 
\begin{eqnarray}
\tilde{\chi}^{+}_i(p_{3}) &\to&
\tilde{\chi}^0_k(p_{5})+\ell^{+}(p_{6})+\nu(p_{7}),\label{eq:p2}\\ 
\mbox{and} & &\nonumber\\
\tilde{\chi}^{-}_j(p_{4}) &\to&
\tilde{\chi}^0_l(p_{8})+\ell^{-}(p_{9})+\bar{\nu}(p_{10})\label{eq:p3}
\end{eqnarray}
The corresponding Feynman diagrams are shown in Fig.~1. 
The production process contains contributions  from $\gamma$- and $Z^0$-exchange
 in the
direct channel and from $\tilde{\nu}$-exchange in the crossed
 channel.
The decay process gets contributions from  
$W^{\pm}$, $\tilde{e}_L$, and $\tilde\nu$ exchange in the different channels.

\noindent From the interaction Lagrangian of the MSSM ( in our notation
 and conventions we follow closely \cite{2haber}):
\begin{eqnarray}
& & {\cal L}_{\gamma \tilde{\chi}_i^{+} \tilde{\chi}_j^{+}} =
- e A_{\mu} \bar{\tilde{\chi}}_i^{+} \gamma^{\mu} \tilde{\chi}_j^{+}
\delta_{ij},\quad e>0, \\
& & {\cal L}_{Z^0\tilde{\chi}_i^{+}\tilde{\chi}_j^{+}} =
 \frac{g}{\cos\theta_W}Z_{\mu}\bar{\tilde{\chi}}_i^{+}\gamma^{\mu}
[O_{ij}^{'L} P_L+O_{ij}^{'R} P_R]\tilde{\chi}_j^{+},\\
& & {\cal L}_{\ell \tilde{\nu}\tilde{\chi}_i^{+}} =
- g U_{i1}^{*} \bar{\tilde{\chi}}_i^{+} P_{L} \nu \tilde{\ell}_L^*
- g V_{i1}^{*} \bar{\tilde{\chi}}_i^{+C} P_L \ell 
 \tilde{\nu}^{*}+\mbox{h.c.},\\
& & {\cal L}_{W^{-}\tilde{\chi}_i^{+}\tilde{\chi}^{0}_k} =
 g W_{\mu}^{-}\bar{\tilde{\chi}}^{0}_k\gamma^{\mu}
[O_{ki}^L P_L+O_{ki}^R P_R]\tilde{\chi}_i^{+} +
\mbox{h.c.},\\
& & {\cal L}_{\ell \tilde{\ell}\tilde{\chi}^{0}_k} =
 g f_{\ell k}^{L} \bar{\ell} P_R \tilde{\chi}^0_k \tilde{\ell}_L
+ g f_{\ell k}^{R} \bar{\ell} P_L \tilde{\chi}^0_k 
\tilde{\ell}_R+\mbox{h.c.},\\
& & {\cal L}_{\nu \tilde{\nu}\tilde{\chi}^{0}_k} =
 g f_{\nu k}^{L} \bar{\nu} P_R \tilde{\chi}^0_k \tilde{\nu}_L+
\mbox{h.c.},
\end{eqnarray}
one gets the couplings:
\begin{eqnarray}
& & L_{\ell}=T_{3\ell}-e_{\ell}\sin^2\theta_W, \quad
 R_{\ell}=-e_{\ell}\sin^2\theta_W,\\
& &\nonumber\\
& & f_{\ell k}^L = -\sqrt{2}\bigg[\frac{1}{\cos
\theta_W}(T_{3\ell}-e_{\ell}\sin^2\theta_W)N_{k2}+
e_{\ell}\sin \theta_W N_{k1}\bigg],\nonumber\\
& & f_{\ell k}^R = -\sqrt{2}e_{\ell} \sin \theta_W\Big[\tan 
\theta_W N_{k2}^*-N_{k1}^*\Big],\nonumber\\
& & f_{\nu k}^L = -\sqrt{2}\frac{1}{\cos\theta_W} 
T_{3\nu}N_{k2},\\
& &\nonumber\\
& & O_{ij}^{'L}=-V_{i1} V_{j1}^{*}-\frac{1}{2} V_{i2} V_{j2}^{*}+
\delta_{ij} \sin^2\Theta_W,\nonumber\\
& & O_{ij}^{'R}=-U_{i1}^{*} U_{j1}-\frac{1}{2} U_{i2}^{*} U_{j2}+
\delta_{ij} \sin^2\Theta_W,\\
& & O_{ki}^L=-1/\sqrt{2}\Big( \cos\beta N_{k4}-\sin\beta N_{k3}
\Big)V_{i2}^{*}
+\Big( \sin\Theta_W N_{k1}+\cos\Theta_W N_{k2} \Big) 
V_{i1}^{*},\nonumber\\
& & O_{ki}^R=+1/\sqrt{2}\Big( \sin\beta N^{*}_{k4}+\cos\beta
N^{*}_{k3}\Big) U_{i2}
+\Big( \sin\Theta_W N^{*}_{k1}+\cos\Theta_W N^{*}_{k2} \Big) 
U_{i1},
\end{eqnarray}
with $i,j=1,2$ and $k=1,\ldots,4$. Here 
$P_{L, R}=\frac{1}{2}(1\mp \gamma_5)$, $g$ is the weak coupling
constant ($g=e/\sin\theta_W$), and $e_\ell$ and 
$T_{3 \ell}$ denote the
charge and the third component of the weak isospin of the 
lepton $\ell$. Furthermore, $\tan\beta=\frac{v_2}{v_1}$ where 
 $v_{1,2}$ are the vacuum expectation values of the two 
neutral Higgs fields,
and $N_{mn}$ are the elements of the unitary $4\times 4$ matrix which 
diagonalizes
the neutral gaugino-higgsino mass matrix in the basis 
$\tilde{\gamma},
\tilde{Z}, \tilde{H}^0_a, \tilde{H}^0_b$.
 The chargino mass eigenstates 
$\tilde{\chi}_i^{+}={\chi_i^{+} \choose \chi_i^{-}}$
are defined by $\chi^{+}_i=V_{i1}w^{+}+V_{i2} h^{+}$ and 
$\chi_j^{-}=U_{j1}w^{-}+U_{j2} h^{-}$. Here $w^{\pm}$ and $h^{\pm}$
are the two-component spinor fields of the W-ino and the charged
higgsinos, respectively. Furthermore, $U_{mn}$ and $V_{mn}$ are
the elements of the
unitary $2\times 2$ matrices which diagonalize the chargino mass matrix. For
details see \cite{bartl}.

The helicity amplitudes $T_P^{\lambda_i\lambda_j}(\alpha)$ for
production and $T_{D,\lambda_i}(\alpha), T_{D,\lambda_j}(\alpha)$ for
the decays, corresponding to the Feynman diagrams in
Fig.~1 are:
\begin{eqnarray}
& &T_P^{\lambda_i \lambda_j}(\gamma)=-\Delta(\gamma)
\delta_{ij} 
\bar{v}(p_2) \gamma^{\mu} u(p_1) \bar{u}_{\lambda_i}(p_3) 
\gamma_{\mu} 
v_{\lambda_j}(p_4),\label{10}\\
& &T_P^{\lambda_i \lambda_j}(Z)=-\Delta(Z)
\bar{v}(p_2) \gamma^{\mu} (L_{\ell} P_L+R_{\ell} P_R) u(p_1) 
\bar{u}_{\lambda_i}(p_3)
\gamma_{\mu} (O^{'L}_{ij} P_L +O^{'R}_{ij} P_R)
v_{\lambda_j}(p_4),\nonumber\\
& &\\
& & T_P^{\lambda_i \lambda_j}(\tilde{\nu})=
-\Delta_{ij}(\tilde{\nu})
\bar{v}(p_2) P_R v_{\lambda_i}(p_3) \bar{u}_{\lambda_j}(p_4) 
P_L u(p_1),\\
& & T_{D,\lambda_i}(W^{+})=-\Delta(W)
\bar{u}(p_5)
\gamma^{\mu} (O_{ki}^L P_L+ O_{ki}^R P_R)
u_{\lambda_i}(p_3)
 \bar{u}(p_7) \gamma_{\mu} P_L v(p_6),\\
& & T_{D,\lambda_i}(\tilde{\ell}_L)=-\Delta_i(\tilde{\ell}_L) 
\bar{u}(p_7) P_R u_{\lambda_i}(p_3) \bar{u}(p_5) P_Lv(p_6),\\
& & T_{D,\lambda_i}(\tilde{\nu})=+\Delta_i(\tilde{\nu})
\bar{u}(p_6) P_L u_{\lambda_i}(p_3) \bar{u}(p_5) P_R v(p_7),\\
& & T_{D,\lambda_j}(W^{-})=-\Delta(W) 
\bar{u}(p_8)\gamma^{\mu} (O_{lj}^{L*} P_R+ O_{lj}^{R*} P_L)
u_{\lambda_j}(p_4) \bar{u}(p_{10}) \gamma_{\mu} P_R v(p_9),\\
& & T_{D,\lambda_j}(\tilde{\ell}_L)=-\Delta_j(\tilde{\ell}_L) 
\bar{u}(p_{10}) P_L u_{\lambda_j}(p_4) \bar{u}(p_8) P_R v(p_9),\\
& & T_{D,\lambda_j}(\tilde{\nu})=+\Delta_j(\tilde{\nu})
\bar{u}(p_9) P_R u_{\lambda_j}(p_4) \bar{u}(p_8) P_L v(p_{10})\label{18}.
\end{eqnarray}
In the following the indices of the couplings $O^{'L,R}_{ij}$ and 
$O^{L,R}_{ki}, O^{L,R}_{lj}$ are suppressed.
 In eqs.~(\ref{10})-(\ref{18}) 
the propagators $\Delta(\gamma)$ etc. include all couplings apart from
$O^{L,R}$ and $O^{'L,R}$. They are defined as follows:
\begin{eqnarray}
& &\Delta(\gamma)=\frac{i e^2}{k^2},\quad
\Delta(Z)=\frac{g^2}{\cos\Theta_W^2}
\frac{i}{k^2-m^2_Z+im_Z\Gamma_Z},
\quad\Delta_{ij}(\tilde{\nu})=
\frac{i g^2 V_{i1}V_{j1}^{*}}{k^2-m^2_{\tilde{\nu}}
+im_{\tilde{\nu}}\Gamma_{\tilde{\nu}}}\nonumber\\
& &\Delta(W)=\frac{g^2}{\sqrt{2}}
\frac{i}{k^2-m^2_W+im_W\Gamma_W},\nonumber\\ 
& &\Delta_i(\tilde{\ell}_L)=\frac{i g^2 U_{i1} f^{*L}_{\ell k} }
{k^2-m^2_{\tilde{\ell}_L}+im_{\tilde{\ell}_L}\Gamma_{\tilde{\ell}_L}},
\quad
\Delta_i(\tilde{\nu})=\frac{i g^2 V_{i1}^{*} f^L_{\nu k}}
{k^2-m^2_{\tilde{\nu}}+im_{\tilde{\nu}}\Gamma_{\tilde{\nu}}}\nonumber\\
& & \Delta_j(\tilde{\ell}_L)=\frac{i g^2 U^{*}_{j1}f^{L}_{\ell l}}
{k^2-m^2_{\tilde{\ell}_L}+im_{\tilde{\ell}_L}\Gamma_{\tilde{\ell}_L}},\quad
\Delta_j(\tilde{\nu})=\frac{i g^2 V_{j1} f^{*L}_{\nu l} }
{k^2-m^2_{\tilde{\nu}}+im_{\tilde{\nu}}\Gamma_{\tilde{\nu}}}\label{23}.
\end{eqnarray}
Here $k^2$ denotes the four-momentum squared of the respective particle.

For the calculation of the amplitude of the combined processes of production and decays, eqs.(\ref{eq:p1}) --
(\ref{eq:p3}), we use the same formalism that we already adopted for
the analogous production
of neutralinos and their decays with polarized beams \cite{moor,3moor}
following the method of \cite{haber}.
 The amplitude for the whole process is
\begin{equation}
T=\Delta(\tilde{\chi}^{+}_i) \Delta(\tilde{\chi}^{-}_j) \sum_{\lambda_i, \lambda_j}T_P^{\lambda_i \lambda_j} 
T_{D, \lambda_i \lambda_j},
\end{equation}
 with the helicity amplitude for the production process 
$T_P^{\lambda_i \lambda_j}$ and that for the decay processes 
 $T_{D,\lambda_i \lambda_j}=T_{D, \lambda_i} T_{D, \lambda_j}$,
 and the propagators $\Delta(\tilde{\chi}^{\pm}_{i,j})=1/[s_{i,j}-m_{i,j}^2+i
m_{i,j}\Gamma_{i,j}]$.
Here $\lambda_{i,j}, s_{i,j}$, $m_{i,j}$, $\Gamma_{i,j}$ 
denote the helicity, four--momentum squared, mass and width of
$\tilde{\chi}^{\pm}_{i,j}$. The amplitude squared
\begin{equation}
|T|^2=|\Delta(\tilde{\chi}^{+}_i)|^2 |\Delta(\tilde{\chi}^{-}_j)|^2 
\rho^{P,\lambda_i\lambda_j \lambda_i'\lambda_j'}
\rho^{D}_{\lambda_i'\lambda_i}\rho^{D}_{\lambda_j'\lambda_j}
\quad\mbox{(sum convention used)}\label{N}
\end{equation}
 is thus composed of the
(unnormalized) spin density production matrix
\begin{equation}
 \rho^{P,\lambda_i\lambda_j \lambda_i'\lambda_j'}=
T_P^{\lambda_i \lambda_j} T_P^{\lambda_i'\lambda_j' *}
\end{equation}
of $\tilde{\chi}^{\pm}_{i,j}$ and
 the decay matrices
\begin{equation}
\rho^{D}_{\lambda_i'\lambda_i}=
T_{D,\lambda_i} T_{D,\lambda_i'}^{*} \quad\mbox{ and}\quad  
\rho^{D}_{\lambda_j'\lambda_j}=
T_{D,\lambda_j} T_{D,\lambda_j'}^{*}.
\end{equation} 
 Interference terms between
various helicity amplitudes preclude factorization in a 
production factor  
$\sum_{\lambda_i \lambda_j} |T_P^{\lambda_i\lambda_j}|^2$ 
times a decay factor
$\overline{\sum}_{\lambda_i \lambda_j} 
|T_{D,\lambda_i\lambda_j}|^2$. 

\noindent The differential cross section in the laboratory system 
is then given by:
\begin{equation}
d\sigma=\frac{1}{8 E_b^2}|T|^2 (2\pi)^4
\delta^4(p_1+p_2-\sum_{i} p_i) d{\rm lips}(p_3\ldots p_{10}),
\end{equation}
$E_b$ denotes the beam energy and $d{\rm lips}(p_3,\ldots,p_{10})=
\prod_{i} \frac{d^3p_i}{(2\pi)^3 2 p_i^0} \quad,i=3,\dots,10$.
\subsection{Spin density production matrix}
We use the general formalism to calculate
 the helicity amplitudes for production
 and decay of a particle
with four-momentum $p$ and mass $m$.
For this purpose we introduce three spacelike four-vectors
$s^a_{\mu},  (a=1,2,3)$, the spin vectors, which together with $p/m$ form an 
orthonormal set \cite{haber}:
\begin{eqnarray}
\frac{p}{m} \cdot s^{a} & = & 0, \\
s^{a} \cdot s^{a'} & = & -{\delta}^{aa'}, \\
s^{a}_{\mu}\cdot s^{a}_{\nu} & = &
-g_{\mu\nu}+\frac{p_{\mu}p_{\nu}}{m^2},\quad
\mbox{summed over a}\label{21}.
\label{321}
\end{eqnarray}
In computing the density matrices for production and
decay one makes
 use of the Bouchiat-Michel formulae \cite{haber}:
\begin{eqnarray}
u_{\lambda'}(p)\bar{u}_{\lambda}(p)&=&\frac{1}{2}
 [\delta_{\lambda{\lambda '}}+\gamma_5 \sq^{a}
  \sigma^{a}_{\lambda{\lambda '}}](\pq+m), \label{322}\\
v_{\lambda'}(p)\bar{v}_{\lambda}(p)&=&\frac{1}{2}
 [\delta_{\lambda'{\lambda}}+\gamma_5 \sq^{a}
 \sigma^{a}_{{\lambda '}\lambda}](\pq-m). \label{323}
\end{eqnarray}
In the amplitude squared, eq.~(\ref{N}), the spin vector $s^{a}$
 enters linearly the  matrices
$\rho^{P,\lambda_i \lambda_j \lambda'_i \lambda'_j}$, 
$\rho^{D}_{\lambda'_i \lambda_i}$ and $\rho^{D}_{\lambda'_j \lambda_j}$.
It is appropriate to separate the spin density production matrix 
 $\rho^{P, \lambda_i \lambda_j \lambda'_i \lambda'_j}$ 
into four parts:
\begin{equation}
\rho^{P,\lambda_i \lambda_j \lambda'_i \lambda'_j}=
\sum_{\alpha \beta}\Big( 
P^{\lambda_i \lambda_j \lambda'_i \lambda'_j}(\alpha\beta)
+\Sigma^{P,\lambda_i \lambda_j \lambda'_i \lambda'_j}_{a}(\alpha\beta)
+\Sigma^{P,\lambda_i \lambda_j \lambda'_i \lambda'_j}_{b}(\alpha\beta)
+\Sigma^{P,\lambda_i \lambda_j \lambda'_i \lambda'_j}
_{ab}(\alpha\beta)\Big),\label{24}
\end{equation}
with $(\alpha\beta)=(\gamma\gamma),(\gamma Z^0),
(\gamma\tilde{\nu}),(Z^0 Z^0), (Z^0 \tilde{\nu}),
(\tilde{\nu} \tilde{\nu})$ denoting the exchanged particles, where
 the second argument corresponds to the complex conjugated amplitude.
One has only these six
combinations of exchanged particles in eq.(\ref{24}) due to the relation: 
\begin{equation}
\rho^{P,\lambda_i \lambda_j \lambda'_i \lambda'_j}
(\alpha \beta)=
(\rho^{P,\lambda_i \lambda_j \lambda'_i \lambda'_j}
(\beta \alpha))^{*}. \label{26}
\end{equation}

\noindent The four parts are defined as follows:
\begin{eqnarray}
& & P^{\lambda_i \lambda_j \lambda'_i \lambda'_j}
(\alpha \beta)=\delta_{\lambda_i \lambda'_i}
\delta_{\lambda_j \lambda'_j} P(\alpha \beta),\label{P}\\
& &\mbox{which is independent of the 
polarization of the charginos},\nonumber\\
& & \Sigma^{P,\lambda_i\lambda_j \lambda'_i\lambda'_j}
_{a}(\alpha \beta)=\delta_{\lambda_j \lambda'_j}\sigma
^{a}_{\lambda_i \lambda'_i}\Sigma^P_{a}(\alpha\beta),\label{Sa}\\ 
& &\mbox{where only the polarization vector $s^a$ of
the chargino $\tilde{\chi}^{+}_i(p_3)$ contributes},\nonumber\\
& & \Sigma^{P,\lambda_i\lambda'_i \lambda_j\lambda'_j}_{b}
(\alpha \beta)=\delta_{\lambda_i\lambda'_i}
\sigma^{b}_{\lambda_j\lambda'_j}\Sigma^P_{b}(\alpha\beta),\label{Sb}\\ 
& &\mbox{where only the polarization vector $s^b$ of 
the chargino  $\tilde{\chi}^{-}_j(p_4)$ contributes},\nonumber\\
& & \Sigma^{P\lambda_i\lambda_j \lambda'_i \lambda'_j}_{ab}
(\alpha \beta)=\sigma^{a}_{\lambda_i\lambda'_i}
\sigma^{b}_{\lambda_j\lambda'_j}
\Sigma^P_{ab}(\alpha\beta),\label{Sab}\\ 
& &\mbox{which includes the polarization vectors $s^a$ and $s^b$ of 
both charginos $\tilde{\chi}^{+}_i$ and 
$\tilde{\chi}^{-}_j$.} \nonumber 
\end{eqnarray}

\vspace*{-.3cm}
\noindent$\sigma^{a}_{\lambda_i\lambda'_i}, \sigma^{b}_{\lambda_j\lambda'_j}$ 
are the $2\times2$ Pauli matrices.

\begin{sloppypar}
In the following we give the analytical formulae for the various
quantities in eq.(\ref{24}). 
Scalar products are
denoted by $(p_i p_j)$. We use the abbrevation  
$[p_i p_j p_k p_l]=i\epsilon_{\mu \nu \rho \sigma} p_i^{\mu} p_j^{\nu}
p_k^{\rho} p_l^{\sigma}$, where the totally antisymmetric tensor
$\epsilon_{\mu \nu \rho \sigma}=+1$, if $\{\mu \nu \rho \sigma\}$ is an
even permutation of $\{0,1,2,3\}$.  
\end{sloppypar} 
It is useful to define  
a coupling $c^{L(R)}(\alpha)$ for the
exchanged particle $(\alpha)$ with
\begin{eqnarray}
& &c^{L}(\gamma)=1\\
& &c^{L}(Z^0)=L_{\ell}\\
& &c^{L}(\tilde{\nu})=1\\
& &c^{R}(\gamma)=1\\
& &c^{R}(Z^0)=R_{\ell}\\
& &c^{R}(\tilde{\nu})=0
\end{eqnarray}
We further introduce the combination  
$c^{P}_{\pm}(\alpha\beta)$ which combines the coupling $c^{L(R)}(\alpha)$
 with the longitudinal beam 
polarization $P_1^3$ and $P_2^3$ of the incoming particles
$e^{-}(p_1)$ and $e^{+}(p_2)$:
\begin{equation}
c^{P}_{\pm}(\alpha\beta)
=\pm c^{L}(\alpha)  c^{L}(\beta)(1-P^3_1)(1+P_2^3)
+c^{R}(\alpha) c^{R}(\beta) (1+P^3_1)(1-P_2^3)\label{34}
\end{equation}
For unpolarized beams one has 
$P^3_1=0=P^3_2$ in
eq.(\ref{34}).
\subsubsection{$P(\alpha\beta)$}
This is the part of eq.(\ref{24}) which is independent of the 
chargino polarization vectors. In $P(\alpha\beta)$ 
 only three different pairs of scalar products contribute: 
\begin{eqnarray}
f_1 & = & (p_1 p_4)(p_2 p_3),\\
f_2 & = & (p_1 p_3)(p_2 p_4),\\
f_3 & = & m_i m_j (p_1 p_2).
\end{eqnarray}
The analytical expressions for $P(\alpha\beta)$, eq.~(\ref{P}),  
 read:
\begin{eqnarray}
P(\gamma \gamma)&=&|\Delta_(\gamma)|^2 
c^{P}_{+}(\gamma \gamma)
\delta_{ij}(f_1+f_2+f_3),\\
P(\gamma Z)&=&2 Re\Big\{\Delta(\gamma)\Delta(Z)^{*}\delta_{ij}
\frac{1}{2} \Big[c^{P}_{+}(\gamma Z)(O^{'L*}+O^{'R*}) (f_1+f_2+f_3)\nonumber\\
& & +c^{P}_{-}(\gamma Z)(O^{'R*}-O^{'L*})(f_1-f_2) \Big]\Big\},\\
P(\gamma \tilde{\nu}) &=& 2 Re\big\{\Delta(\gamma) 
\Delta_{ij}^{*}(\tilde{\nu})\delta_{ij}\frac{1}{4}
c^{P}_{+}(\gamma \tilde{\nu}) (2 f_1 +f_3)\big\},\\
P(ZZ)&=&|\Delta(Z)|^2\frac{1}{2}
\Big[c^{P}_{+}(Z Z)\Big((|O^{'L}|^2+|O^{'R}|^2) (f_1+f_2)\nonumber\\ 
   & & -(O^L O^{'R*}+O^R O^{'L*})(-f_3)\Big)
+c^{P}_{-}(Z Z)(|O^{'R}|^2-|O^{'L}|^2)(f_1-f_2)\Big],\\
P(Z \tilde{\nu})&=&2 Re\big\{\Delta(Z) \Delta_{ij}^{*}(\tilde{\nu}) 
\frac{1}{4} c^{P}_{+}(Z\tilde{\nu})(2 O^{'L} f_1+O^{'R}
f_3)\big\},\\
P(\tilde{\nu} \tilde{\nu})&=& |\Delta_{ij}(\tilde{\nu})|^2
\frac{1}{4} c^{P}_{+}(\tilde{\nu} \tilde{\nu})f_1.
\end{eqnarray}
\subsubsection{$\Sigma^P_{a}(\alpha\beta)$ and 
 $\Sigma^P_{b}(\alpha\beta)$}
This is the part of eq.(\ref{24}) which contains only one polarization
vector, either $s^a$ of $\tilde{\chi}^{+}_i$ or $s^b$ of $\tilde{\chi}^{-}_j$.
In $\Sigma^{P}_{a}(\alpha \beta)$, eq.(\ref{Sa}), the following five
 products  with the polarization vector $s^a$ appear:
\begin{eqnarray}
f^a_1&=& m_i (p_2 p_4)(p_1 s^a),\label{44}\\
f^a_2&=& m_i (p_1 p_4)(p_2 s^a),\\
f^a_3&=& m_j (p_2 p_3)(p_1 s^a),\\
f^a_4&=& m_j (p_1 p_3)(p_2 s^a),\label{47}\\
f^a_5&=& m_j [p_2 p_1 s^a p_3].\label{48}
\end{eqnarray}
The contributions $\sum_a^P(\alpha\beta)$ in eq.(\ref{Sa}) are:
\begin{eqnarray}
\Sigma^P_a(\gamma \gamma)&=&
|\Delta(\gamma)|^2 c^{P}_{-}(\gamma\gamma)
\delta_{ij}(-f^a_1+f^a_2-f^a_3+f^a_4),\label{49}\\
\Sigma^P_a(\gamma Z) &=&2 Re\Big\{\Delta(\gamma)\Delta^{*}(Z)
\frac{1}{2}\delta_{ij}
\Big[c^{P}_{+}(\gamma Z)(O^{'R*}-O^{'L*})
(f^a_1+f^a_2)+c^{P}_{-}(\gamma Z)
\nonumber\\
& &\Big( (O^{'R*}+O^{'L*}) (-f^a_1+f^a_2-f^a_3+f^a_4)
     -(O^{'R*}-O^{'L*}) f^a_5\Big)\Big]\Big\},\\
\Sigma^P_a(\gamma \tilde{\nu}) &=& 2 Re\big\{\Delta(\gamma) 
\Delta_{ij}^{*}(\tilde{\nu}) \frac{1}{4} \delta_{ij}
c^{P}_{+}(\gamma\tilde{\nu}) (-2f^a_2+f^a_3-f^a_4-f^a_5)\big\},\\
\Sigma^P_a(ZZ)&=&|\Delta(Z)|^2 \frac{1}{2}
\Big[c^{P}_{+}(ZZ) (|O^{'R}|^2-|O^{'L}|^2) (f^a_1+f^a_2)
\nonumber\\
& & + c^{P}_{-}(ZZ)\Big((O^{'L} O^{'R*}+O^{'R} O^{'L*}) (-f^a_3+f^a_4)
\nonumber\\
& &     +(|O^{'R}|^2+|O^{'L}|^2) (-f^a_1+f^a_2)
       -(O^{'L} O^{'R*}-O^{'R} O^{'L*})f^a_5\Big)\Big],\\
\Sigma^P_a(Z \tilde{\nu})&=&2 Re\big\{\Delta(Z) \Delta_{ij}^{*}(\tilde{\nu}) 
\frac{1}{4}c^{P}_{+}(Z \tilde{\nu})
\big(-2 O^{'L} f^a_2-O^{'R} (-f^a_3+f^a_4+f^a_5)\big)\big\},\\
\Sigma^P_a(\tilde{\nu} \tilde{\nu})&=& |\Delta_{ij}(\tilde{\nu})|^2
\frac{1}{4} c^{P}_{+}(\tilde{\nu} \tilde{\nu}) (-f^a_2).\label{54}
\end{eqnarray}
In the analogous formulae for $\sum_b^P(\alpha\beta)$, eq.~(\ref{Sb}), the 
following five products containing the polarization vector $s^b$ contribute:
\begin{eqnarray}
f^b_1&=& m_i (p_2 p_4)(p_1 s^b),\label{55}\\
f^b_2&=& m_i (p_1 p_4)(p_2 s^b),\\
f^b_3&=& m_j (p_2 p_3)(p_1 s^b),\\
f^b_4&=& m_j (p_1 p_3)(p_2 s^b),\label{58}\\
f^b_5&=& m_i [p_2 p_1 s^b p_4],\label{59}
\end{eqnarray}
The corresponding formulae for $\sum_b^P(\alpha\beta)$ are obtained by
substituting in eqs.(\ref{49})-(\ref{54}):
\begin{equation}
f^a_1\to -f^b_4,\quad f^a_2\to -f^b_3,\quad f^a_3\to -f^b_2,\quad
f^a_4\to -f^b_1,\quad f^a_5\to -f^b_5.
\end{equation}
\subsubsection{$\Sigma^P_{ab}(\alpha\beta)$}
This is the part of eq.(\ref{24}) which contains  
both polarization vectors $s^a$ of $\tilde{\chi}^{+}_i$ and   
$s^b$ of $\tilde{\chi}^{-}_j$. It can be expressed by  
 eight different combinations of products with 
both spin vectors:
\begin{eqnarray}
f^{ab}_1&=&(p_3 p_4)(p_1 s^a)(p_2 s^b),\\
f^{ab}_2&=&m_i m_j (p_1 s^a)(p_2 s^b),\\
f^{ab}_3&=&(p_3 p_4)(p_1 s^b)(p_2 s^a),\\
f^{ab}_4&=&m_i m_j (p_1 s^b)(p_2 s^a),\\
f^{ab}_5&=&(s^a s^b)
          [(p_1 p_4)(p_2 p_3)-(p_1 p_2)(p_3 p_4)+(p_1 p_3)(p_2 p_4)],
\label{77}\\
f^{ab}_6&=&(p_3 s^b)
          [(p_1 p_2)(p_4 s^a)-(p_1 p_4)(p_2 s^a)-(p_2 p_4)(p_1 s^a)],\\
f^{ab}_7&=&(p_4 s^a)[(p_1 p_3)(p_2 s^b)+(p_2 p_3)(p_1 s^b)],\\
f^{ab}_8&=&(p_2 p_4)[s^b s^a p_3 p_1]-(p_3 p_1)[s^b s^a p_2 p_4]
            +(p_2 s^b)[s^a p_3 p_1 p_4]+(s^a p_1)[s^b p_3 p_2 p_4]\label{78}.
\end{eqnarray}
The contributions $\sum_{ab}^P(\alpha\beta)$ in eq.(\ref{Sab}) are:
\begin{eqnarray}
\Sigma^P_{ab}(\gamma \gamma)&=&|\Delta(\gamma)|^2 
 c^{P}_{+}(\gamma\gamma)
\delta_{ij}(-f^{ab}_1-f^{ab}_2-f^{ab}_3-f^{ab}_4-f^{ab}_5
-f^{ab}_6+f^{ab}_7)\\
\Sigma^P_{ab}(\gamma Z)&=&2 Re\Big\{\Delta(\gamma)\Delta^{*}(Z)
\frac{1}{2}\delta_{ij}
\Big[c^{P}_{+}(\gamma Z)
\Big( (O^{'R*}-O^{'L*}) f^{ab}_8
\nonumber\\
& &
+(O^{'R*}+O^{'L*}) 
(-f^{ab}_1-f^{ab}_2-f^{ab}_3-f^{ab}_4-f^{ab}_5-f^{ab}_6
+f^{ab}_7)\Big)\nonumber\\
& &
+c^{P}_{-}(\gamma Z)(O^{'R*}-O^{'L*}) (f^{ab}_2-f^{ab}_4)\Big]\Big\},\\
\Sigma^P_{ab}(\gamma \tilde{\nu})&=&2 Re\big\{\Delta(\gamma) 
\Delta_{ij}^{*}(\tilde{\nu}) 
\frac{1}{4} \delta_{ij}
c^{P}_{+}(\gamma\tilde{\nu}) (-f^{ab}_1-f^{ab}_3-2f^{ab}_4
-f^{ab}_5-f^{ab}_6+f^{ab}_7-f^{ab}_8)\big\},\nonumber\\
& &\\
\Sigma^P_{ab}(ZZ)&=&|\Delta(Z)|^2 \frac{1}{2}
\Big[c^{P}_{+}(ZZ)
   \Big((O^{'L} O^{'R*}+O^{'R} O^{'L*})\nonumber\\
& &
(-f^{ab}_1-f^{ab}_3-f^{ab}_5-f^{ab}_6+f^{ab}_7)
+(|O^{'R}|^2+|O^{'L}|^2) (-f^{ab}_2-f^{ab}_4)
\nonumber\\
& &
-(O^{'L} O^{'R*}-O^{'R} O^{'L*})
 (-f^{ab}_8)
\Big)\nonumber\\ 
& &
+c^{P}_{-}(ZZ)
(|O^{'L}|^2-|O^{'R}|^2) (-f^{ab}_2+f^{ab}_4)\Big]\\
\Sigma^P_{ab}(Z \tilde{\nu})
&=&2 Re\big\{\Delta(Z) \Delta_{ij}^{*}(\tilde{\nu})  \frac{1}{4}
c^{P}_{+}(Z\tilde{\nu})
\big(-2 O^{'L} f^{ab}_4\nonumber\\
& &
-O^{'R}(f^{ab}_1+f^{ab}_3+f^{ab}_5+f^{ab}_6-f^{ab}_7+f^{ab}_8)\big)\big\}\\
\Sigma^P_{ab}(\tilde{\nu} \tilde{\nu})&=&
|\Delta_{ij}(\tilde{\nu})|^2 \frac{1}{4} 
c^{P}_{+}(\tilde{\nu}\tilde{\nu})(- f^{ab}_4)
\end{eqnarray}
\subsection{Decay matrix}
In the following we give the analytical formulae for the
(unnormalized) decay matrices for both decays
$\tilde{\chi}_i^{+}(p_3)\to\tilde{\chi}_k^0(p_5) 
+\ell^{+}(p_6)+\nu(p_7)$ and 
$\tilde{\chi}_j^{-}(p_4)\to\tilde{\chi}_l^0(p_8)
+\ell^{-}(p_9)+\bar{\nu}(p_{10})$ (see Fig.~1).
In the following $m_k$ and $m_l$ denote the masses of the neutralinos 
$\tilde{\chi}^0_k$ and $\tilde{\chi}^0_l$. 
 Analogously to eq.(\ref{24}) the decay matrices
$\rho^D_{\lambda'_i\lambda_i}$
and $\rho^D_{\lambda'_j\lambda_j}$
can both be separated into two parts:
\begin{eqnarray}
\rho^D_{\lambda'_i\lambda_i}&=&\sum_{\alpha \beta} 
\Big(D_{\lambda'_i\lambda_i}(\alpha \beta)+
\Sigma^D_{a,\lambda'_i\lambda_i}(\alpha \beta)\Big),\label{24a}\\
\rho^D_{\lambda'_j\lambda_j}&=&\sum_{\alpha \beta} 
\Big(D_{\lambda'_j\lambda_j}(\alpha \beta)+
\Sigma^D_{b,\lambda'_j\lambda_j}(\alpha \beta)\Big),\label{24b}
\end{eqnarray}
with $(\alpha \beta)=(WW),(W\tilde{\ell}_L),(W\tilde{\nu}),
(\tilde{\ell}_L\tilde{\ell}_L),(\tilde{\ell}_L\tilde{\nu}),
(\tilde{\nu}\tilde{\nu})$ denoting the exchanged particle. The second
argument corresponds to the complex conjugated amplitude. Here we have
also used the relation analogous to eq.(\ref{26}). The combinations 
$(\beta\alpha)$
are therefore already included.

\noindent The two parts in eqs.(\ref{24a}) and (\ref{24b}) are given 
as follows:
\begin{eqnarray}
& & D_{\lambda'_i\lambda_i}(\alpha
  \beta)=\delta_{\lambda'_i\lambda_i}D_i(\alpha\beta),\label{Di}\\
& & \mbox{which
is independent of the polarization vector of the chargino
  $\tilde{\chi}^{+}_i(p_3)$,}\nonumber\\
& & \Sigma^D_{a,\lambda'_i\lambda_i}(\alpha\beta)
=\sigma^a_{\lambda'_i\lambda_i}\Sigma_a^D(\alpha\beta),\label{Si}\\
& & \mbox{where the polarization vector $s^a$ of the chargino
$\tilde{\chi}^{+}_i(p_3)$ contributes},\nonumber\\
& & D_{\lambda'_j\lambda_j}(\alpha
  \beta)=\delta_{\lambda'_j\lambda_j}D_j(\alpha\beta),\label{Dj}\\
& & \mbox{which
is independent of the polarization vector of the chargino
  $\tilde{\chi}^{-}_j(p_4)$},\nonumber\\
& & \Sigma^D_{b,\lambda'_j\lambda_j}(\alpha\beta)
=\sigma^b_{\lambda'_j\lambda_j}\Sigma_b^D(\alpha\beta),\label{Sj}\\
& &\mbox{where the polarization vector $s^b$ of the chargino
$\tilde{\chi}^{-}_j(p_4)$ contributes.}\nonumber
\end{eqnarray}
\subsubsection{$D_i(\alpha\beta)$ and $D_j(\alpha\beta)$}
These are the parts of eq.(\ref{24a}) and eq.(\ref{24b}) which are  
independent of the polarization vector $s^a$ ($s^b$) of the decaying
chargino $\tilde{\chi}_i^{+}$ ($\tilde{\chi}^{-}_j$).
In the term $D_i(\alpha\beta)$, eq.(\ref{Di}), for the decay of
$\tilde{\chi}^{+}_i(p_3)$ only three different pairs of scalar products 
contribute:
\begin{eqnarray}
g_1&=&(p_5 p_7) (p_3 p_6), \\
g_2&=&(p_5 p_6) (p_3 p_7), \\
g_3&=&(m_i m_k) (p_6 p_7).
\end{eqnarray}
The analytical expressions for $D_i(\alpha\beta)$, eq.(\ref{Di}),  read: 
\begin{eqnarray}
D_i(W^{+} W^{+})&=&|\Delta(W)|^2 4
\Big[ (|O^L|^2+|O^R|^2) (g_1+g_2)
-(O^{L*} O^R +O^L O^{R*}) 
g_3\nonumber\\
& &\phantom{|\Delta(W)|^2 2}
-(|O^R|^2-|O^L|^2)(g_1-g_2) \Big],\\
D_i(W^{+} \tilde{\ell}_L)&=&2 Re\Big\{\Delta(W) \Delta_i^{*}(\tilde{\ell}_L)2
\Big(2O^R g_2 -O^L g_3\Big)\Big\},\\
D_i(W^{+} \tilde{\nu})&=&2 Re\Big\{-\Delta(W) \Delta_i^{*}(\tilde{\nu})2 
\Big(2 O^L g_1-O^R g_3\Big)\Big\},\\
D_i(\tilde{\ell}_L \tilde{\ell}_L)&=&|\Delta_i(\tilde{\ell}_L)|^2 2 g_2,\\
D_i(\tilde{\ell}_L \tilde{\nu})&=&2 Re \big\{\Delta_i(\tilde{\ell}_L)
\Delta_i^{*}(\tilde{\nu}) g_3\big\},\\
D_i(\tilde{\nu} \tilde{\nu})&=&|\Delta_i(\tilde{\nu})|^2 2 g_1.
\end{eqnarray}
The corresponding scalar products for the decay of the 
 $\tilde{\chi}^{-}_j(p_4)$ and the expressions for $D_j(\alpha\beta)$,
 eq.(\ref{Dj}),
 are obtained by the following substitutions:
\begin{eqnarray}
& & p_5 \to p_8, p_6 \to p_9, p_7 \to p_{10}, \quad
m_i \to m_j, m_k \to m_l \label{O1}\\
& & O^L_{ki}\to O^{L*}_{lj}, O^R_{ki}\to O^{R*}_{lj}\label{O3}\\
& & \Delta_i(\tilde{\ell}_L)\to \Delta_j(\tilde{\ell}_L),\quad 
    \Delta_i(\tilde{\nu}) \to \Delta_j(\tilde{\nu}) \quad\quad 
\mbox{(see eq.(\ref{23}))}\label{O2}.
\end{eqnarray}
\subsubsection{$\Sigma_a^D(\alpha\beta)$ and $\Sigma_b^D(\alpha\beta)$}
These are the parts of eq.(\ref{24a}) and eq.(\ref{24b}) which contain 
the polarization vector either $s^a$ of $\tilde{\chi}^{+}_i$ or $s^b$
of $\tilde{\chi}^{-}_j$. 
In the term $\Sigma^D_a(\alpha\beta)$ of eq.(\ref{Si}) 
 the following five
 products with the polarization vector $s^a$ appear:
\begin{eqnarray}
g^a_1&=&m_i (p_5 p_7) (p_6 s^a), \label{96}\\
g^a_2&=&m_i (p_5 p_6) (p_7 s^a), \\
g^a_3&=&m_k (p_3 p_7) (p_6 s^a), \\
g^a_4&=&m_k (p_3 p_6) (p_7 s^a), \label{98}\\
g^a_5&=&m_k [s^a p_3 p_7 p_6].\label{99}
\end{eqnarray}
\begin{eqnarray}
\Sigma^D_a(W^{+} W^{+})&=&|\Delta(W)|^2 4
\Big[(|O^R|^2-|O^L|^2) (g^a_1+g^a_2)
-(O^{L*} O^R +O^L O^{R*})(-g^a_3+g^a_4)\nonumber\\ 
& &
-(|O^R|^2+|O^L|^2) (g^a_1-g^a_2) 
    +(O^{L*} O^R -O^L O^{R*}) g^a_5\Big],\\
\Sigma^D_a(W^{+} \tilde{\ell}_L)&=&2 Re\Big\{
\Delta(W) \Delta_i^{*}(\tilde{\ell}_L)2
\Big[2 O^R g^a_2 -O^L (-g^a_3+g^a_4+g^a_5)\Big]\Big\},\\
\Sigma^D_a(W^{+} \tilde{\nu})&=&2 Re \Big\{
-\Delta(W) \Delta_i^{*}(\tilde{\nu}) 2 
\Big[-2 O^L g^a_1-O^R (-g^a_3+g^a_4-g^a_5)\Big]\Big\},\\
\Sigma^D_a(\tilde{\ell}_L \tilde{\ell}_L)&=&
|\Delta_i(\tilde{\ell}_L)|^2 2 g^a_2,\\
\Sigma^D_a(\tilde{\ell}_L \tilde{\nu})&=&2 Re\big\{-\Delta_i(\tilde{\ell}_L)
\Delta^{*}_i(\tilde{\nu}) (g^a_3-g^a_4+g^a_5)\big\},\\
\Sigma^D_a(\tilde{\nu} \tilde{\nu})&=&|\Delta_i(\tilde{\nu})|^2 2 (-g^a_1)
\end{eqnarray}
The corresponding expressions $\Sigma^D_b(\alpha\beta)$, 
eq.(\ref{Sj}), for the 
decay of $\tilde{\chi}^{-}_j(p_4)$ one obtains by  
the same substitutions as in 2.3.1, eqs.~(\ref{O1})-(\ref{O2}), 
 and the additional substitution $s^a \to
 -s^b$ in eqs.(\ref{96})-(\ref{98}) and $[s^a p_3 p_7
 p_6]\to [(-s^b) p_4 p_{10} p_9]^*$ in eq.(\ref{99}).
\subsection{Amplitude squared for production and decay}
The amplitude squared $|T|^2$ of the combined processes of production and decays, eqs.(\ref{eq:p1}) -- (\ref{eq:p3}),  is given by: 
\begin{eqnarray}
& &|T|^2=\sum_{(\alpha\beta)(\alpha_i\beta_i)(\alpha_j\beta_j)}
         4\Big(P(\alpha \beta) D_i(\alpha_i\beta_i) D_j(\alpha_j\beta_j)
    +\Sigma^P_a(\alpha \beta) \Sigma^D_a(\alpha_i\beta_i) 
D_j(\alpha_j\beta_j)\nonumber\\
& &\phantom{|T|^2=\sum_{\alpha \beta \alpha_i\beta_i\alpha_j\beta_j}}
    +\Sigma^P_b(\alpha \beta) \Sigma^D_b(\alpha_j\beta_j)
D_i(\alpha_i\beta_i) 
    +\Sigma^P_{ab}(\alpha \beta) \Sigma_a^D(\alpha_i\beta_i) 
\Sigma_b^D(\alpha_j\beta_j)\Big).\label{95}
\end{eqnarray}
The arguments label the six combinations of the exchanged particles:
\begin{eqnarray}
& & (\alpha\beta):(\gamma\gamma),(\gamma Z),
(\gamma\tilde{\nu}),
(ZZ),(Z\tilde{\nu}),(\tilde{\nu}\tilde{\nu}),\\
& & (\alpha_i\beta_i):(W^{+}W^{+}),(W^{+}\tilde{\ell}_L),
(W^{+}\tilde{\nu}),
(\tilde{\ell}_L\tilde{\ell}_L),(\tilde{\ell}_L\tilde{\nu}),
(\tilde{\nu}\tilde{\nu}),\\
& & (\alpha_j\beta_j):(W^{-}W^{-}),(W^{-}\tilde{\ell}_L),
(W^{-}\tilde{\nu}),
(\tilde{\ell}_L\tilde{\ell}_L),(\tilde{\ell}_L\tilde{\nu}),
(\tilde{\nu}\tilde{\nu}).
\end{eqnarray}
The first product in eq.(\ref{95}) 
is the part obtained by neglecting all spin correlations between 
production and decay. The second and third term describe the correlations between
the production and the decay process either of
$\tilde{\chi}_i^+ \to \tilde{\chi}^0_k \ell^+ \nu_\ell$ or 
$\tilde{\chi}^-_j \to
\tilde{\chi}^0_\ell \ell^- \bar{\nu}_{\ell}$, and in the last term  
correlations between both decay
processes are included. 
\begin{itemize}
\item In the first term of eq.(\ref{95}) 
only scalar products appear which can be expressed 
by the Mandelstam variables $s, t, u$ 
 for the production and decay processes.
\item To obtain the second (third) term of eq.(\ref{95})
one has to calculate all combinations of 
$f^a_m \times g^a_n, m,n=1,\ldots,5$ ($f^b_m \times g^b_n, m,n=1,\ldots,5$)
using eq.(\ref{21}). In this way one gets additional scalar products:
\begin{equation}
(p_{1,2}p_{6,7}), \quad\quad (p_{1,2}p_{9,10}),
\end{equation}
describing correlations between production and
decay. These scalar products can not be expressed by Mandelstam
variables. They contain the angle between the incoming electron and
the outgoing lepton in the laboratory system.
 The combinations $[s^{(a,b)} p_k p_l p_m]$, 
eqs.(\ref{48}), (\ref{59}), (\ref{99}), 
are due to complex parameters and to the term in the propagators,
eq.(\ref{23}), containing the width of the exchanged
particle. They lead to triple product correlations (using
eq.(\ref{21})):
\begin{equation}
[p_{1,2} p_3 p_7 p_6], \quad [p_{1,2} p_4 p_{10} p_9],\quad
[p_{6,7} p_2 p_1 p_3], \quad [p_{9,10} p_2 p_1 p_4].
\end{equation} 
\item To obtain the last term of eq.(\ref{95}) one has to calculate all combinations of
$f^{ab}_m \times g^a_{n_i} \times g^b_{n_j},  m=1,\dots,8,n_i,n_j=1,\dots,5$ 
using again eq.(\ref{21}).
Then due to the combinations $(s^a s^b)$, eq.(\ref{77}), and 
$[s^a s^b p_k p_l]$, eq.(\ref{78}),
 also correlations between the decay products of both charginos appear: 
\begin{equation}
(p_{6,7} p_{9,10}),\quad
[p_{6,7} p_4 p_{10} p_9], \quad [p_{9,10} p_3 p_7 p_6],\quad  
[p_{9,10} p_{6,7} p_3 p_1], \quad [p_{9,10} p_{6,7} p_2 p_4].
\end{equation}
\end{itemize}

If only the decay of one chargino, e.g. $\tilde{\chi}^-_j$ is
considered, one has to sum over the spin of $\tilde{\chi}^+_i$ so that
 in  eq.(\ref{95}) 
$D_i(\alpha_i\beta_i)=1$ and 
$\Sigma^D_a(\alpha_i\beta_i)=0$. 
\section{Numerical Results and Discussion}
\vspace{-.3cm}
In the MSSM \cite{2haber} the masses and couplings of charginos and neutralinos
 are determined by the  parameters $M' , M, \mu,
\tan\beta$, with $M'$ usually fixed by the GUT relation 
$M'=\frac{5}{3} M \tan^2\Theta_W$.
Since we do not consider CP violation in the following analysis, 
the parameters and the couplings 
 of charginos and neutralinos can be chosen real. The neutralino 
and chargino mass mixing matrices can be found in \cite{2haber,fraas1}. 

The masses of the sleptons and of the sneutrinos are related to the
parameters $M$ and $\tan\beta$ of the MSSM and to the common scalar
mass $m_0$ at the unification point by the renormalization group
equations \cite{hall}:
\begin{equation}
m^2_{\tilde{f}_L}=m_0^2 + 0.79 M^2 + m_Z^2 \cos2\beta (T_3^f-Q_f
\sin^2\Theta_W).
\end{equation}
Here $T_3^f$ and $Q_f$ denote the third component of the weak isospin
and the charge of the corresponding fermion.

In the following numerical analysis we study the pair
production of the lighter chargino $\tilde{\chi}^{\pm}_1$ with
unpolarized beams, where one of the charginos decays leptonically, 
$\tilde{\chi}^{-}_1
\to \tilde{\chi}^0_1+\ell^{-}+\bar{\nu}_{\ell}$. 
In order to illustrate the influence
of the gaugino--higgsino mixing of 
$\tilde{\chi}^\pm_1$ and $\tilde{\chi}^0_1$ we consider
four representative scenarios.

In Table~1 we give the parameters and the mass eigenvalues (including
their sign). For the scalar mass parameter we choose in general
$m_0=200$~GeV. Since the angular distribution depends also
 on the value of $m_0$, we compare in scenario~(B) the numerical
results for $m_0=200$~GeV with those for $m_0=100$~GeV. In Table~2 the
gaugino and higgsino components of the chargino and of $\tilde{\chi}^0_1$ are
given.

In scenarios~(A) and (B) for low and high $\tan\beta$, respectively, 
$\tilde{\chi}^\pm_1$ is almost a pure W-ino and $\tilde{\chi}^0_1$ is
almost a pure B-ino. In scenarios~(C) and (D) for low and high 
$\tan\beta$, respectively,
both $\tilde{\chi}^\pm_1$ and $\tilde{\chi}^0_1$ have
dominating higgsino components.

In the following we will discuss the angular  
distribution  $d\sigma/d\cos\Theta_{-}$ (where $\Theta_{-}$ is the
angle between the outgoing
 $\ell^{-}$ and the electron beam) and the
forward-backward asymmetry
\begin{equation} 
A_{FB}=\frac{\sigma(\cos\Theta_{-}>0)-\sigma(\cos\Theta_{-}<0)}
{\sigma(\cos\Theta_{-}>0)+\sigma(\cos\Theta_{-}<0)}
\label{afb}\end{equation}
of the lepton $\ell^{-}$
from the leptonic decay 
$\tilde{\chi}^{-}_1 \to \tilde{\chi}^0_1+e^{-}+\bar{\nu}$ for
$\sqrt{s}=192$~GeV and $\sqrt{s}=200$~GeV. 

The total cross sections are given in Table~3 for $\sqrt{s}=192$~GeV
and in Table~4 for $\sqrt{s}=200$~GeV. 
They are independent of the spin correlations and
factorize into the chargino production cross section times the 
leptonic branching ratio of $\tilde{\chi}^{-}_1$ \cite{tata}. 
For comparison the different angular
distributions in Figs.~4,~5 and 6 are normalized to
the total cross section.

To demonstrate the significance of spin correlations we compare in
Figs.~2 and 3 the angular distributions for $\tan\beta=40$ and 
$\sqrt{s}=200$~GeV with the results one obtains by factorizing the process
in production  and decay. The spin effect is sizeable for the
gaugino-like charginos of scenario~(B), Fig.~2. It is largest in the
forward and in the backward direction. The forward-backward asymmetry
is $A_{FB}=+39.5\%$, one order of magnitude larger as that obtained neglecting
the spin correlations between production and decay. For the
higgsino-like scenario~(D), Fig.~3, the influence of spin correlations is
less significant. 
 Quite generally, the influence of spin correlations is  much more pronounced for gaugino-like
charginos than for higgsino-like ones. 

Especially for
gaugino-like charginos the spin effects  depend sensitively on the value
of $\tan\beta$. For $\tan\beta=3$ they are smaller than for
$\tan\beta=40$. For scenario~(A) with $\tan\beta=3$ 
the forward-backward asymmetry 
is $A_{FB}=+6.3\%$ ($A_{FB}=7.9\%$) with spin correlations and
$A_{FB}=2.0\%$ ($A_{FB}=3.8\%$) without spin correlations
 at $\sqrt{s}=192$~GeV ($\sqrt{s}=200$~GeV).
 In Fig.~4 we compare the angular distributions for gaugino-like charginos for
$\sqrt{s}=192$~GeV and $m_0=200$~GeV and two values of $\tan\beta$, 
$\tan\beta=3$, scenario~(A), and $\tan\beta=40$, scenario~(B). For both
values of $\tan\beta$ the forward hemisphere is favoured, and  
the forward-backward asymmetry $A_{FB}$ is +6.3\% for $\tan\beta=3$ and
+38.9\% for $\tan\beta=40$. 

In contrast, for the higgsino-like charginos in scenario~(C) and (D),
Fig.~5, the distributions are rather flat with the backward direction
somewhat favoured. The forward-backward asymmetry is very small:
$A_{FB}=-1.6\%$ for $\tan\beta=3$, and $A_{FB}=-3.4\%$ for $\tan\beta=40$.
  The shape of the angular distribution is similar at 
$\sqrt{s}=200$~GeV with the same values of $A_{FB}$ as for $\sqrt{s}=192$~GeV 
(see Table~4). 

In the
case of gaugino-like charginos the size of the 
forward-backward asymmetry
depends on the value of $m_0$. In Fig.~6 we 
therefore compare for scenario~(B) the
angular distributions for
$m_0=100$~GeV and $m_0=200$~GeV. For $m_0=200$~GeV the 
forward-backward asymmetry is
$A_{FB}=+39.5\%$ and for $m_0=100$~GeV the asymmetry is $A_{FB}=+29.1\%$.

\section{Summary and conclusions} We have calculated the analytical expression for the
differential cross section for 
$e^{+}e^{-} \to \tilde{\chi}^{+}_i \tilde{\chi}^{-}_j$ with polarized beams and the
subsequent leptonic decays $\tilde{\chi}^{+}_{i} \to \tilde{\chi}^0_k
\ell^{+}\nu_\ell$ and $\tilde{\chi}^{-}_{j} \to \tilde{\chi}^0_l \ell^{-}\bar{\nu}_\ell$,
taking into account the complete spin correlations between production and decays. The
differential cross section is composed of the joint spin-density matrix of the two
charginos and the decay matrices for their leptonic decays. The corresponding expressions also include spin correlations between the leptons coming from the decays of both charginos. Moreover, the analytical formulae 
for the production and the decay matrices can be used as building blocks for processes
with other chargino decay channels or for processes with other chargino production
mechanisms, as e.g. $e^{-}\gamma$ or $\gamma\gamma$ collisions. 

In the numerical analysis we have studied 
the pair production of the lighter chargino
$\tilde{\chi}^{\pm}_1$ with unpolarized beams and to the
leptonic decay of one of them, $\tilde{\chi}^{-}_1 \to
\tilde{\chi}^0_1+\ell^{-}+
\bar{\nu}$. We have calculated the angular distribution of $\ell^-$ in the laboratory frame at $\sqrt{s} =
192$~GeV and $\sqrt{s} = 200$~GeV for four representative MSSM
scenarios which differ in the mixing character of the chargino and 
in the parameter $\tan\beta$.
For the case of a gaugino-like $\tilde{\chi}^{-}_1$ we have also studied the influence of
the scalar mass parameter $m_0$ on the shape of the angular distribution. Generally, the
effect of spin correlations is much more significant for gaugino-like charginos  than
for higgsino-like ones. 

The shape of the lepton angular distribution and the size of the
forward-backward asymmetry depend on the nature of the charginos. For
higgsino-like charginos the angular distribution is nearly flat with a
 small forward-backward asymmetry, whereas for gaugino-like ones
the forward-backward asymmetry is between 6.3\% and 39.5\% for
$m_0=200$~GeV. In the case of gaugino-like
charginos, however,  
the shape of the angular distribution and the forward-backward asymmetry depend
on the scalar mass $m_0$. 

In conclusion, we have found that for a precise analysis of the lepton angular
distributions in the decays of charginos produced in $e^+e^- \to 
\tilde{\chi}^+_i
\tilde{\chi}^-_j$ the inclusion of spin correlations between
production and decay is necessary for gaugino-like charginos.

\section{Acknowledgement}
We thank H.~Wachter for checking the analytical formulae 
and we are grateful to V.~Latussek for his support in the development of the
numerical program. G.M.-P. was supported by {\it
  Friedrich-Ebert-Stiftung}. This work was also supported by 
the German Federal Ministry for
Research and Technology (BMBF) under contract number
05 7WZ91P (0), by the Deutsche Forschungsgemeinschaft under
contract Fr 1064/2-2 and the `Fond zur
F\"orderung der wissenschaftlichen Forschung' of Austria, Project
No. P10843-PHY.

\begin{table}[h]
\begin{center}
\begin{tabular}{|l||c|c|c|c|c|c|c|c|}
  & M & $\mu$ & $\tan\beta$ & $m_0$ &  
$m_{\tilde{\chi}^0_1}$ &
 $m_{\tilde{\chi}^{\pm}_1}$ 
& $m_{\tilde{e}_{L}}$ & $m_{\tilde{\nu}}$
 \\ \hline\hline
A & 87  & $-800$ & 3   & 200 & 45 & 91 & 219  
 & 207 
\\ \hline
B & 91.5 & $-800$ & 40 & 200 (100) & 45 & 91 & 221 (137) & 206 (112)
\\ \hline
C & 363  & 107 & 3 & 200 & 76 & $-91$ & 382 & 375
 \\ \hline
D & 302 & 99 & 40 & 200  & 75 & $-91$ 
& 338 & 328
 \\ \hline
\end{tabular}
\caption{Parameters M, $\mu, \tan\beta$, 
and $m_0$, and the resulting mass eigenvalues. 
All masses are given in [GeV].}
\end{center}
\begin{center}
\begin{tabular}{|l|cc|cc|cccc|}
 & \multicolumn{2}{c|}{$\tilde{\chi}^{+}_1$}
 & \multicolumn{2}{c|}{$\tilde{\chi}^{-}_1$} 
 & \multicolumn{4}{c|}{$\tilde{\chi}^0_1$}\\ \hline
& ($w^{+}|$&$H^{+}$) & 
($w^{-}|$&$H^{-}$) 
& ($\tilde{\gamma}|$&$\tilde{Z}|$&$\tilde{H}^0_a|$&$\tilde{H}^0_b$)\\ \hline
A &($+1.0|$&$+.03$)&($+.99|$&$+.13$)&($+.90|$&$-.44|$&$-.03|$&-.04)
 \\ \hline
B &($+1.0|$&$-.01$)&($+.99|$&$+.14$)&($+.87|$&$-.48|$&$+.0004|$&$-.05$)
 \\ \hline
C &($+.33|$&$-.95$)&($-.18|$&$+.98$)&($+.18|$&$-.35|$&$+.78|$&$+.48$)
 \\ \hline
D &($+.38|$&$-.92$)&($-.13|$&$+.99$)&($-.19|$&$+.34|$&$-.58|$&$-.72$)
\\ \hline
\end{tabular}
\caption{Mixing character of the chargino 
$\tilde{\chi}^\pm_{1}$ and the
 neutralino $\tilde{\chi}^0_1$.}
\end{center}
\begin{center}
\begin{tabular}{|l||c|c|c|c|}
  & A & B & C & D  \\ \hline\hline
$\sigma_t$ /fb & 339  & 587 & 177 & 175 \\ \hline
$A_{FB}$ & $+.063$ & $+.389$ & $-.016$ & $-.034$  
 \\ \hline
\end{tabular}
\caption{$\sigma(e^{-}e^{+}\to \tilde{\chi}^{+}_1
  \tilde{\chi}^{-}_1)\times$BR($\tilde{\chi}^{-}_1\to\tilde{\chi}^0_1
\ell^{-}\bar{\nu}$) 
and forward-backward asymmetries $A_{FB}$, eq.(\ref{afb}), for
  $\sqrt{s}=192$~GeV.}
\end{center}
\begin{center}
\begin{tabular}{|l||c|c|c|c|c|}
  & A & B, $m_0=200$~GeV & B, $m_0=100$~GeV & C & D  \\ \hline\hline
$\sigma_t$ /fb & 368  & 638 &  177  & 204 & 202 \\ \hline
$A_{FB}$ & $+.079$ & $+.395$ & $.291$ & $-.016$ & $-.034$ 
 \\ \hline
\end{tabular}
\caption{$\sigma(e^{-}e^{+}\to \tilde{\chi}^{+}_1
  \tilde{\chi}^{-}_1)\times$BR($\tilde{\chi}^{-}_1\to\tilde{\chi}^0_1
\ell^{-}\bar{\nu}$) 
and forward-backward asymmetries $A_{FB}$, eq.(\ref{afb}), for
  $\sqrt{s}=200$~GeV.}
\end{center}
\end{table}

\clearpage

\begin{figure}
\hspace{1cm}
\begin{minipage}[t]{3.5cm}
\begin{center}
{\setlength{\unitlength}{1cm}
\begin{picture}(2.5,2.5)
\put(-.6,-1.1){\includegraphics{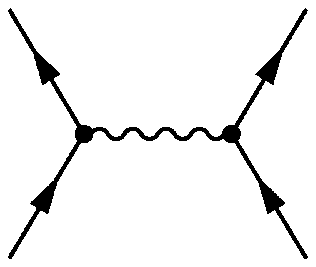}}
\put(-.8,-1.5){$e^{-}(p_1)$}
\put(2,-1.5){$\tilde{\chi}^{-}_j(p_4)$}
\put(-.8,1.5){$e^{+}(p_2)$}
\put(2,1.5){$\tilde{\chi}^{+}_i(p_3)$}
\put(1,.4){$\gamma$}
\end{picture}}
\end{center}
\end{minipage}
\hspace{2cm}
\vspace{.8cm}
\begin{minipage}[t]{3.5cm}
\begin{center}
{\setlength{\unitlength}{1cm}
\begin{picture}(2.5,2.5)
\put(-1.2,-1.1){\includegraphics{prog.ps}}
\put(-1.5,-1.5){$e^{-}(p_1)$}
\put(1.3,-1.5){$\tilde{\chi}^{-}_j(p_4)$}
\put(-1.5,1.5){$e^{+}(p_2)$}
\put(1.3,1.5){$\tilde{\chi}^{+}_i(p_3)$}
\put(.4,.4){$Z^0$}
\end{picture}}
\end{center}
\end{minipage}
\hspace{2cm}
\vspace{.8cm}
\begin{minipage}[t]{3.5cm}
\begin{center}
{\setlength{\unitlength}{1cm}
\begin{picture}(2.5,2)
\put(-1.2,-1.3){\includegraphics{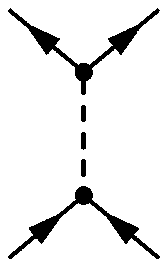}}
\put(-2,-1.5){$e^{-}(p_1)$}
\put(-2,1.5){$e^{+}(p_2)$}
\put(-.1,-1.5){$\tilde{\chi}^{-}_j(p_4)$}
\put(-.1,1.5){$\tilde{\chi}^{+}_i(p_3)$}
\put(-.9,0){$\tilde{\nu}$}
 \end{picture}}
\end{center}
\end{minipage}

\vspace{.5cm}

\begin{minipage}[t]{3.5cm}
\begin{center}
{\setlength{\unitlength}{1cm}
\begin{picture}(5,2.5)
\put(+.8,-.9){\includegraphics{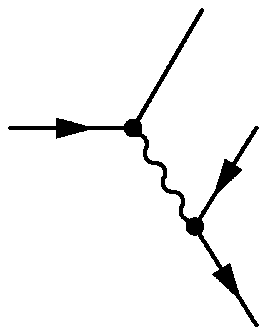}}
\put(3.9,-1.3){$\nu(p_7)$}
\put(.6,1.2){$\tilde{\chi}^{+}_i(p_{3})$}
\put(3.7,1){$\ell^{+}(p_6)$}
\put(3.3,2.2){$\tilde{\chi}^0_k(p_5)$}
\put(1.9,.1){$W^{+}$}
\end{picture}}
\end{center}
\end{minipage}
\hspace{2cm}
\vspace{.8cm}
\begin{minipage}[t]{3.5cm}
\begin{center}
{\setlength{\unitlength}{1cm}
\begin{picture}(5,4.5)
\put(+.3,+1){\includegraphics{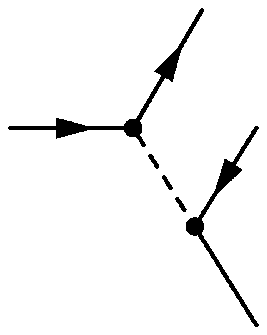}}
\put(2.5,4.2){$\nu (p_7)$}
\put(2.9,3){$\ell^{+}(p_6)$}
\put(-.1,3.2){$\tilde{\chi}^{+}_i(p_3)$}
\put(1.6,.8){$\tilde{\chi}^0_k(p_5)$}
\put(1.5,2){$\tilde{\ell}_{L}$}
\end{picture}}
\end{center}
\end{minipage}
\begin{minipage}[t]{3.5cm}
\begin{center}
{\setlength{\unitlength}{1cm}
\begin{picture}(5,4)
\put(1.5,.5){\includegraphics{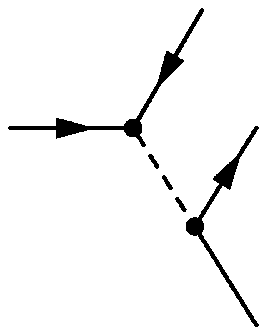}}
\put(4.2,2.5){$\nu (p_7)$}
\put(3.7,3.6){$\ell^{+}(p_6)$}
\put(2.7,.4){$\tilde{\chi}^0_k (p_5)$}
\put(1.1,2.7){$\tilde{\chi}^{+}_i (p_3)$}
\put(2.8,1.5){$\tilde{\nu}$}
\end{picture}}
\end{center}
\end{minipage}

\vspace{-.7cm}

\begin{minipage}[t]{3.5cm}
\begin{center}
{\setlength{\unitlength}{1cm}
\begin{picture}(5,2.5)
\put(.8,-.9){\includegraphics{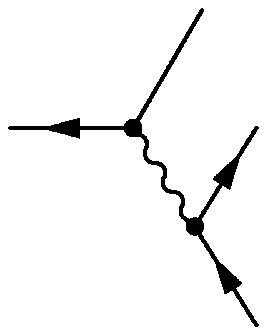}}
\put(3.9,-1.3){$\bar{\nu}(p_{10})$}
\put(.6,1.2){$\tilde{\chi}^{-}_j(p_{4})$}
\put(3.7,1){$\ell^{-}(p_9)$}
\put(3.3,2.2){$\tilde{\chi}^0_l(p_8)$}
\put(1.9,.1){$W^{-}$}
\end{picture}}
\end{center}
\end{minipage}
\hspace{2cm}
\vspace{.8cm}
\begin{minipage}[t]{3.5cm}
\begin{center}
{\setlength{\unitlength}{1cm}
\begin{picture}(5,4.5)
\put(+.3,+1){\includegraphics{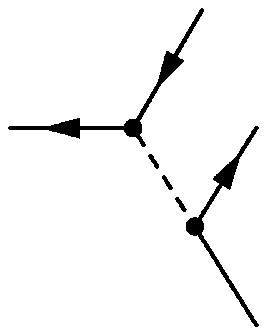}}
\put(2.5,4.2){$\bar{\nu} (p_{10})$}
\put(2.9,3){$\ell^{-}(p_9)$}
\put(-.1,3.2){$\tilde{\chi}^{-}_j(p_4)$}
\put(1.6,.8){$\tilde{\chi}^0_l(p_8)$}
\put(1.5,2){$\tilde{\ell}_{L}$}
\end{picture}}
\end{center}
\end{minipage}
\begin{minipage}[t]{3.5cm}
\begin{center}
{\setlength{\unitlength}{1cm}
\begin{picture}(5,4)
\put(+1.5,.5){\includegraphics{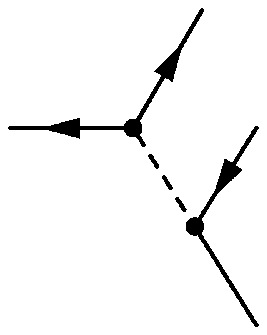}}
\put(4.2,2.5){$\bar{\nu} (p_{10})$}
\put(3.7,3.6){$\ell^{-}(p_9)$}
\put(2.7,.4){$\tilde{\chi}^0_l (p_8)$}
\put(1.1,2.7){$\tilde{\chi}^{-}_j (p_4)$}
\put(2.8,1.5){$\tilde{\nu}$}
\end{picture}}
\end{center}
\end{minipage}
\vspace*{-1cm}
\caption{Feynman graphs for production and leptonic decays of
  charginos. \label{channel}}
\end{figure}

\begin{figure}
\begin{picture}(8,7)
\put(0,0){\includegraphics{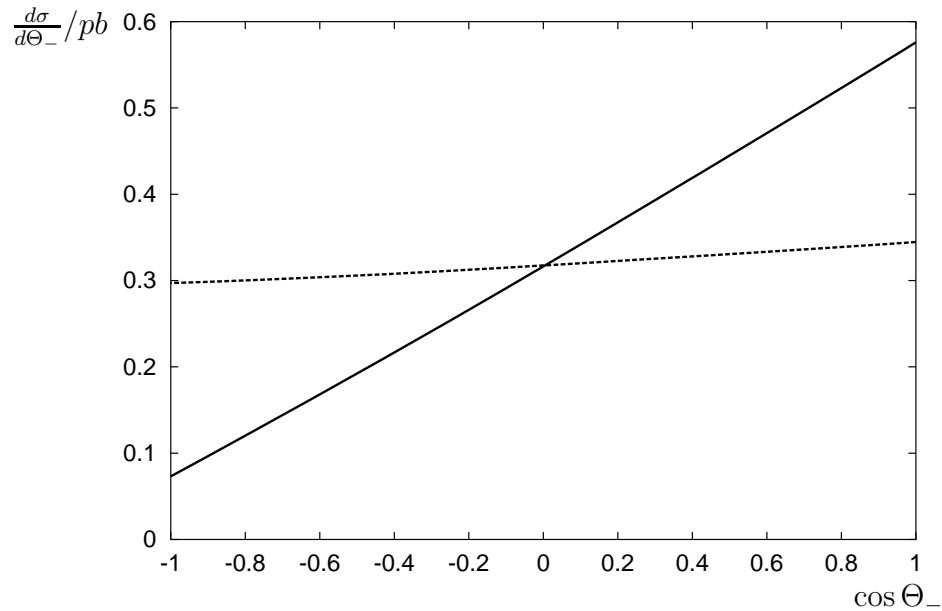}}
\put(12,1.5){$ \cos\Theta_{-}$}
\put(.8,9.1){$ \frac{d\sigma}{d\Theta_{-}}/pb$}
\end{picture}
\vspace*{-1cm}
\caption{Lepton angular distributions in scenario (B) for $m_0$ = 200~GeV at
$\sqrt{s}=200$ GeV with spin correlations 
 (solid) and without spin correlations (dotted).\label{w1}}
\end{figure}

\begin{figure}
\begin{picture}(8,7)
\put(0,0){\includegraphics{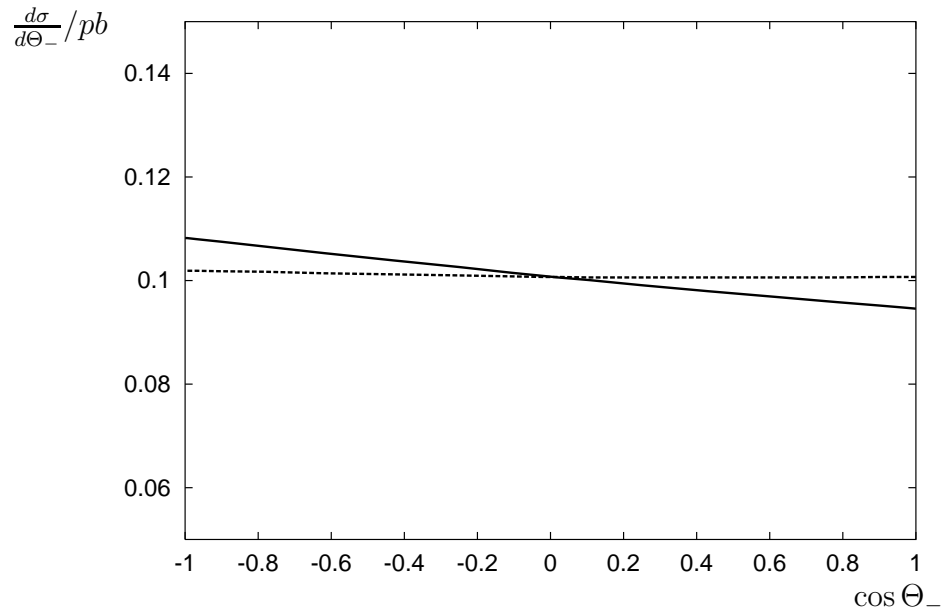}}
\put(12,1.5){$\cos\Theta_{-}$}
\put(0.8,9.1){$ \frac{d\sigma}{d\Theta_{-}}/pb$}
\end{picture}
\vspace*{-1cm}
\caption{Lepton angular distribution in scenario (D)
at $\sqrt{s}=200$ GeV
with spin correlations (solid), and without spin correlations
(dotted). \label{w2}}
\end{figure}

\begin{figure}
\begin{picture}(8,7)
\put(0,0){\includegraphics{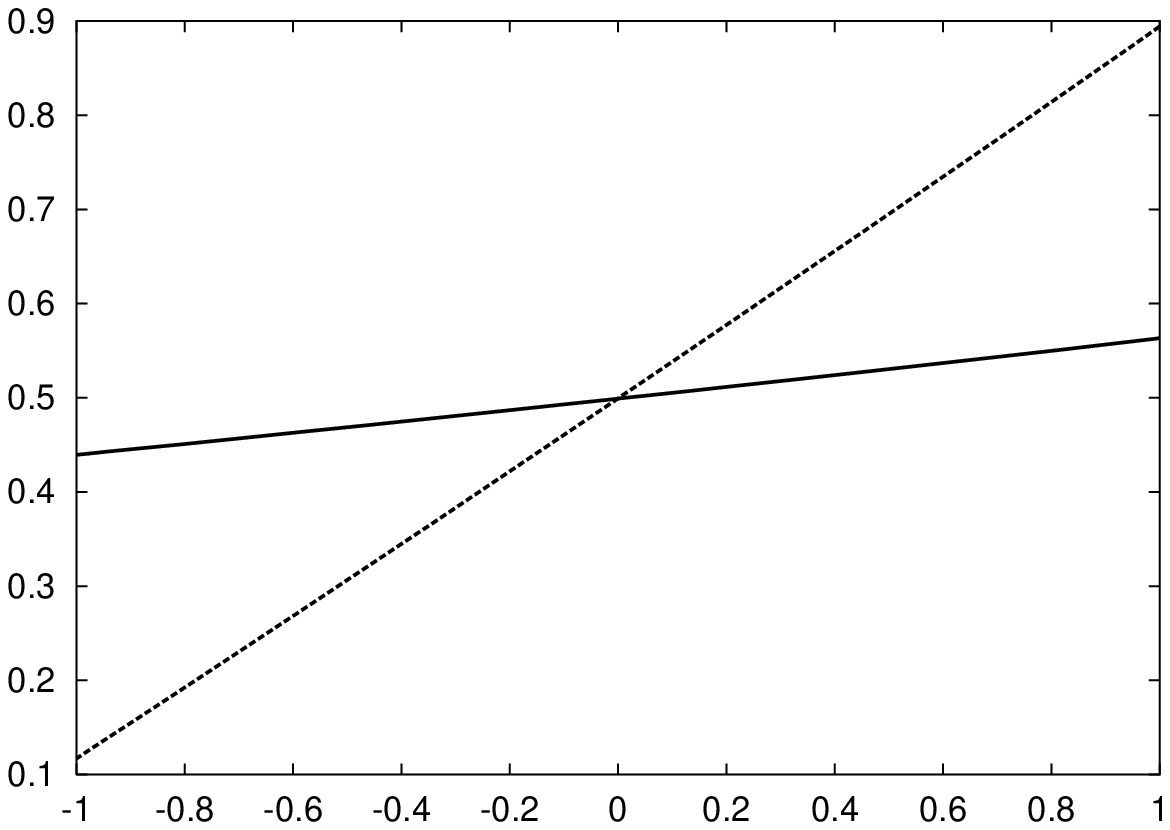}}
\put(12,1.5){$ \cos\Theta_{-}$}
\put(1,9.1){$\frac{1}{\sigma_t}\frac{d\sigma}{d\Theta_{-}}$}
\end{picture}
\vspace*{-1cm}
\caption{Lepton angular distributions in scenario (A) (solid) and scenario (B)
(dotted) for
 $\sqrt{s}=192$ GeV, normalized to the total cross section.\label{w3}}
\end{figure}

\begin{figure}
\begin{picture}(8,7)
\put(0,0){\includegraphics{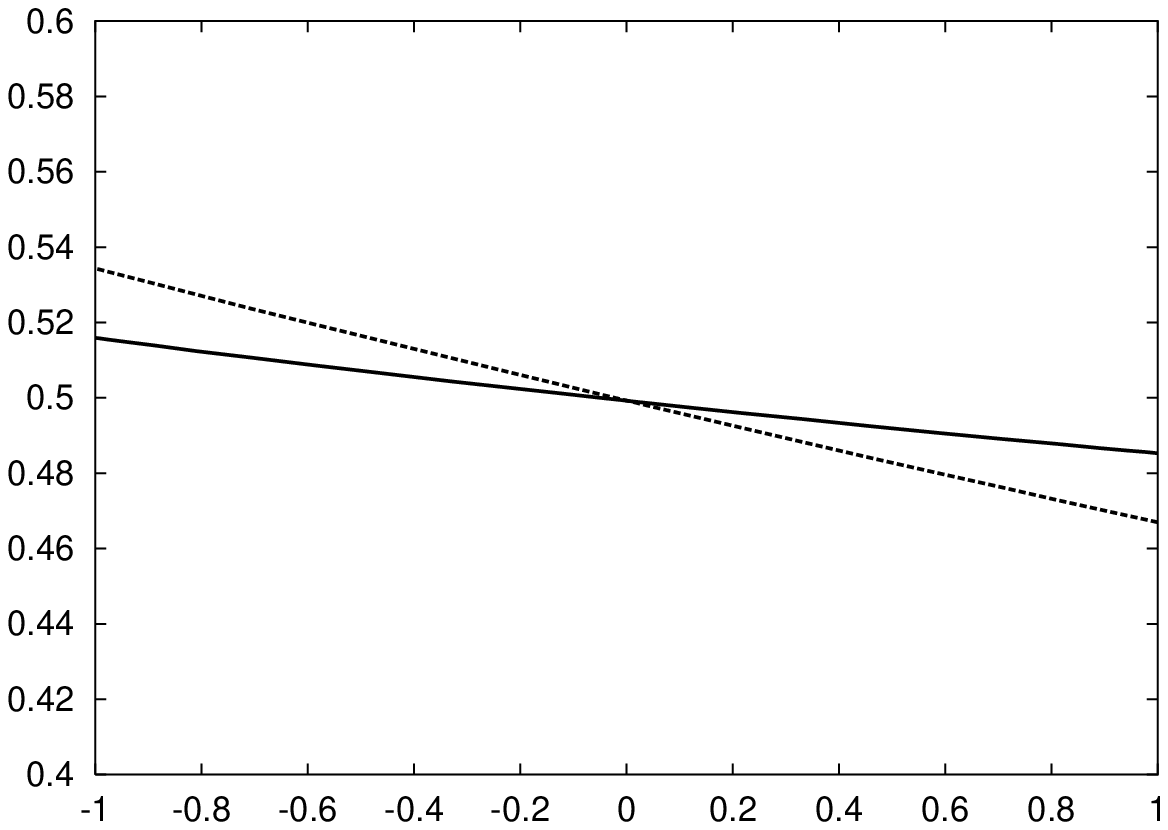}}
\put(12,1.5){$\cos\Theta_{-}$}
\put(1,9.1){$\frac{1}{\sigma_t}\frac{d\sigma}{d\Theta_{-}} $}
\end{picture}
\vspace*{-1cm}
\caption{Lepton angular distribution in scenario (C) (solid) and scenario (D) 
(dotted) at $\sqrt{s}=192$ GeV, normalized to the total cross 
section.\label{w4}}
\end{figure}

\begin{figure}
\begin{picture}(8,7)
\put(0,0){\includegraphics{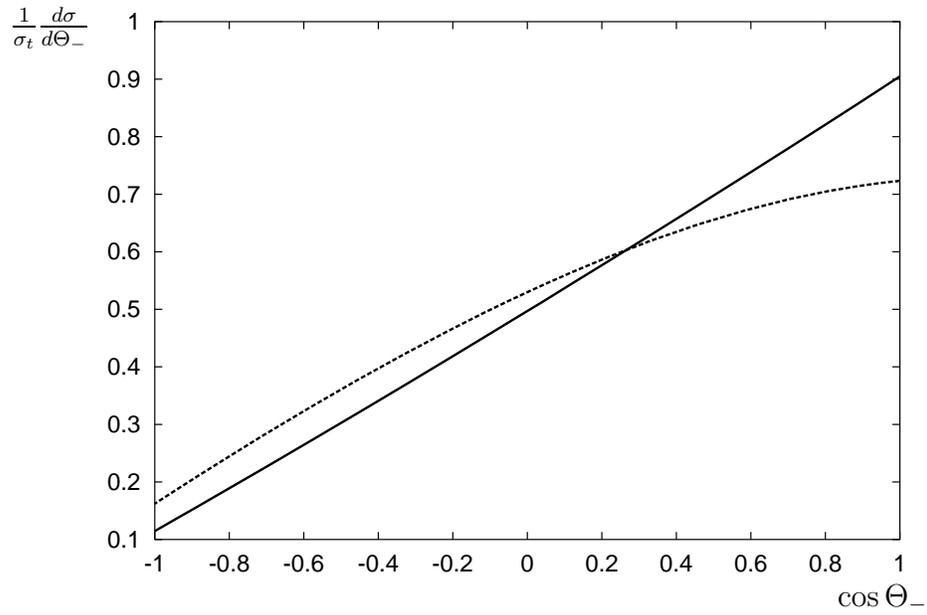}}
\put(12,1.5){$ \cos\Theta_{-}$}
\put(1,9.1){$ \frac{1}{\sigma_t}\frac{d\sigma}{d\Theta_{-}}$}
\end{picture}
\vspace*{-1cm}
\caption{Lepton angular distribution in scenario (B) 
at $\sqrt{s}=200$ GeV with $m_0=200$~GeV (solid) and 
$m_0=100$ GeV (dotted), normalized to the total cross section.\label{w5}}
\end{figure}
\end{document}